\documentclass[nofootinbib,aps,pra,showpacs,superscriptaddress,preprint]{revtex4}
\usepackage{amsmath}
\usepackage{subfig}
\usepackage{graphicx}

\catcode`ð=\active
\defð{\u{g}}
\catcode`Ð=\active
\defÐ{\u{G}}
\catcode`Ý=\active
\defÝ{\. I}
\catcode`ö=\active
\defö{\"{o}}
\catcode`Ö=\active
\defÖ{\"O}
\catcode`ü=\active
\defü{\"{u}}
\catcode`Ü=\active
\defÜ{\"{U}}
\catcode`Þ=\active
\defÞ{\c{S}}
\catcode`þ=\active
\defþ{\c{s}}
\catcode`ý=\active
\defý{{\i}}
\catcode`ç=\active
\defç{\d{c}}
\catcode`Ç=\active
\defÇ{\d{C}}

\begin{document}

\title{Step-Up and Step-Down Operators of a two-term Molecular Potential via Nikiforov-Uvarov Method}
\author{\small Altuð Arda}
\email[E-mail: ]{arda@hacettepe.edu.tr}\affiliation{Department of
Physics Education, Hacettepe University, 06800, Ankara,Turkey}
\author{\small Ramazan Sever}
\email[E-mail: ]{sever@metu.edu.tr}\affiliation{Department of
Physics, Middle East Technical  University, 06531, Ankara,Turkey}

\begin{abstract}
The creation and annihilation operators of a two-term diatomic molecular potential are studied and
it is observed that they satisfy the commutation relations of a SU(1,1) algebra. To study the Lie algebraic realization of the present
potential, the normalized eigenfunctions and eigenvalues are computed by using the Nikiforov-Uvarov method.
\end{abstract}

\maketitle

\newpage

\section{Introduction}
The two-term diatomic molecular potential under consideration [1]
\begin{eqnarray}
V(r)=-V_{0}\,\frac{e^{-\beta r}}{1-qe^{-\beta
r}}+V_{1}\,\frac{e^{-2\beta r}}{(1-qe^{-\beta r})^2}\,,
\end{eqnarray}
has been firstly proposed by Sun to study of some diatomic
molecules. Jia and co-workers have studied the bound states of
this potential for the Schrödinger (SE) equation in the view of
supersymmetric approach [2] and in Ref. [3], the energy spectra of
the potential is obtained in terms of Green's function for the
same domain. Furthermore, Arda \textit{et. al.} have studied the
bound states and the corresponding normalized wave functions of
this potential for the SE equation with any $\ell$-values for the
cases where $q \geq 1$ and $q=0$. In addition, they have also
given the analytical results for the Manning-Rosen potential, the
Hulth\'{e}n potential and the generalized Morse potential as
special cases with their numerical results [4]. As a continuation,
in the present work, we intend to search the creation and
annihilation operators of the two-term molecular potential and
find that the dynamical group satisfied by the ladder operators of
the present potential. For this aim, we obtain firstly the
normalized eigenfunctions of the potential by using the
Nikiforov-Uvarov (NU) method [5] to present the ladder operators.

The group theoretical realizations within the quantum mechanics
have been received great attentions in literature. The
factorization method is used to get exact solutions of some
solvable potentials with the ladder operators [6, 7]. The
perturbed ladder operator method is applied to quantum mechanical
systems with the help of unperturbed eigenvalues and
eigenfunctions [6, 8]. Dong and co-workers have applied a
'different' factorization method to various types of molecules
[9-16].

The organization of the present work is as follows. In Section 2,
we search the normalized wave functions with their energy spectra
of this potential by using NU-method which is briefly given in
Appendix A. In Section 3, we present the raising and lowering
operators of the potential and show that the commutators of the
ladder operators satisfy the Lie algebra of a SU(1,1) group.
\section{Normalized Wave Functions and Energy Spectrum}
The radial Schrödinger equation is written [17]
\begin{eqnarray}
\frac{d^{2}R_{n\ell}(r)}{dr^2}+\left[\frac{2\mu}{\hbar^2}\left[E_{n\ell}-V(r)\right]-\frac{\ell(\ell+1)}{r^2}\right]R_{n\ell}(r)=0\,,
\end{eqnarray}
where $\ell$ is the angular momentum quantum number, $m$ is the
particle mass, $V(r)$ is the central potential and $E_{n\ell}$ is
the non-relativistic energy.

Inserting Eq. (1) into Eq. (2) and using a new variable
$x=qe^{-\beta r }$ ($x \rightarrow q$ for $r \rightarrow 0$, $x
\rightarrow 0$ for $r \rightarrow \infty$) gives the following
equation without the centrifugal term
\begin{eqnarray}
\frac{d^{2}R_{n}(x)}{dx^2}+\frac{1-x}{x(1-x)}\frac{dR_{n}(x)}{dx}+\left[-a^2_{1}-a^2_{2}x-a^2_{3}x^2\right]R_{n}(x)=0\,.
\end{eqnarray}
with the abbreviations
\begin{subequations}
\begin{align}
&a^2_{1}=-\frac{2\mu E_{n}}{\beta^2\hbar^2}\,,\\
&a^2_{2}=\frac{2\mu}{\beta^2\hbar^2}\left(2E_{n}-\frac{V_{0}}{q}\right)\,,\\
&a^2_{3}=\frac{2\mu}{\beta^2\hbar^2}\left(\frac{V_{1}}{q^2}+\frac{V_{0}}{q}-E_{n}\right)\,,
\end{align}
\end{subequations}
where $V_{0}, V_{1}, \beta$ and $q$ are real parameters defined by
$V_{1}=D_{0}(e^{\mu}-q)$, $V_{0}=2V_{1}$, $\beta=\mu/r_{0}$, where
$D_{0}$ is the depth of the potential, $r_{0}$ is the equilibrium
of the molecule and $q$ is the shape parameter.

Comparing Eq. (3) with Eq. (A1)we obtain
\begin{subequations}
\begin{align}
&\tilde{\tau}(x)=1-x\,,\\
&\sigma(x)=x-x^2\,,\\
&\tilde{\sigma}(x)=-a^2_{3}x^2-a^2_{2}x-a^2_{1}\,,
\end{align}
\end{subequations}
with $\sigma''(x)=-2$. Eq. (A7) gives
\begin{eqnarray}
\pi(x)=-\frac{x}{2}\mp
\sqrt{\left(\frac{1}{4}+a^2_{3}-k\right)x^2+\left(a^2_{2}+k\right)x+a^2_{1}\,}\,,
\end{eqnarray}
We obtain the parameter $k$ is determined by setting the
discriminant of the square root in Eq. (6)
\begin{subequations}
\begin{align}
&k_{1}=-a^2_{2}-2a^2_{1}-a_{1}A\,,\\
&k_{2}=-a^2_{2}-2a^2_{1}+a_{1}A\,\,;\,\,\,\,A=\sqrt{1+\frac{8\mu
V_{1}}{q^2\beta^2\hbar^2}\,}\,.
\end{align}
\end{subequations}
From Eqs. (7a) and (7b) we get the followings for $\pi(x)$,
respectively,
\begin{subequations}
\begin{align}
&\pi(x)=a_{1}-\left(a_{1}+\frac{1+A}{2}\right)x\,,\\
&\pi(x)=a_{1}-\left(a_{1}+\frac{1-A}{2}\right)x\,.
\end{align}
\end{subequations}
The parameter $\lambda$ required for the method is obtained from
Eq. (A8) with the help of Eq. 8(a)
\begin{eqnarray}
\lambda=-a^2_{2}-2a^2_{1}-\left(a_{1}+\frac{1}{2}\right)-A\left(a_{1}+\frac{1}{2}\right)\,,
\end{eqnarray}
and the other one $\lambda_{n}$ is written
\begin{eqnarray}
\lambda_{n}=n\left(2+2a_{1}+A\right)+n(n-1)\,,
\end{eqnarray}
where used $\tau(x)=1+2a_{1}-(2+2a_{1}+A)x$ and
$\tau'(x)=-(2+2a_{1}+A)$.

Setting $\lambda=\lambda_{n}$ and using the parameter values given
in Eqs. (4a)-(4c) gives the following energy eigenvalue equation
\begin{eqnarray}
a_{1}+\frac{A}{2}=-\left(n+\frac{1}{2}\right)\mp a_{3}\,,
\end{eqnarray}
and the analytical expression for the energy spectra
\begin{eqnarray}
E_{n}=-\frac{\beta^2\hbar^2}{2\mu}\Bigg\{\frac{\left[2n+1+\sqrt{1+\frac{8\mu
V_{1}}{q^2\beta^2\hbar^2}\,}\right]^2-\frac{8\mu}{\beta^2\hbar^2}\,\left(\frac{V_{0}}{q}+\frac{V_{1}}{q^2}\right)}
{4\left[2n+1+\sqrt{1+\frac{8\mu
V_{1}}{q^2\beta^2\hbar^2}\,}\right]}\Bigg\}^2\,.
\end{eqnarray}

In order to find the wave functions, we first use Eq. (A6) to
present the function $\rho(x)$
\begin{eqnarray}
\rho(x) \sim x^{2a_{1}}(1-x)^{A}\,,
\end{eqnarray}
and inserting it into Eq. (A5) gives
\begin{eqnarray}
\varphi_{n}(x) \sim
\frac{1}{x^{2a_{1}}(1-x)^{A}}\frac{d^n}{dx^{n}}\left[x^{n+2a_{1}}(1-x)^{n+A}\right]\,,
\end{eqnarray}
which means that the second part of the wave function in Eq. (A2)
is written as the Jacobi polynomials [18]
\begin{eqnarray}
\varphi_{n}(x) \sim P_{n}^{(2a_{1},A)}(1-2x)\,,
\end{eqnarray}
Eq. (A4) gives the first part of wave function as
\begin{eqnarray}
\psi(x) \sim x^{a_{1}}(1-x)^{(1+A)/2}\,,
\end{eqnarray}
Finally, the total wave functions are given as
\begin{eqnarray}
R_{n}(x)=A_{n} x^{a_{1}}(1-x)^{(1+A)/2}P_{n}^{(2a_{1},A)}(1-2x)\,.
\end{eqnarray}
where $A_{n}$ is a constant determined from the normalization
condition. Choosing a new variable as $x=qy$, we write the
normalization condition
\begin{eqnarray}
\int_{0}^{1}\left|A_{n}\right|^2q^{2a_{1}}y^{2a_{1}}(1-qy)^{1+A}\left[P_{n}^{(2a_{1},A)}(1-2qy)\right]
\left[P_{m}^{(2a_{1},A)}(1-2qy)\right]dy=1\,,
\end{eqnarray}
Using the following representation of the Jacobi polynomials [18]
\begin{eqnarray}
P_{n}^{(\xi_{1},\,\xi_{2})}(z)=\frac{1}{2^n}\sum_{k=0}^{n}\,\Bigg(\begin{array}{c}
  a+n \\
  k
\end{array}\Bigg)\,\Bigg(\begin{array}{c}
  b+n \\
  n-k
\end{array}\Bigg)\,(1+z)^{k}(1-z)^{n-k}\,,
\end{eqnarray}
where binomial coefficient $\Bigg(\begin{array}{c}
  n \\
  r
\end{array}\Bigg)=\frac{n!}{r!(n-r)!}=\frac{\Gamma(n+1)}{\Gamma(r+1)\Gamma(n-r+1)}$\,.
Hence, Eq. (18) becomes
\begin{eqnarray}
\left|A_{n}\right|^2q^{2a_{1}+n+m-k-\ell+1}(-1)^{n+m-k-\ell}\,[g(n,k)\times
g(m,\ell)]\int_{0}^{1}y^{2a_{1}+n-k+m-\ell}(1-qy)^{1+A+k+\ell}dy=1\,,\nonumber\\
\end{eqnarray}
where $g(n,k)$ and $g(m,\ell)$ are two arbitrary functions of the
parameters $a_{1}$ and $A$ and given by
\begin{subequations}
\begin{align}
&g(n,k)=\sum_{k=0}^{n}\,\Bigg(\begin{array}{c}
  n+2a_{1} \\
  k
\end{array}\Bigg)\,\Bigg(\begin{array}{c}
  n+A \\
  n-k
\end{array}\Bigg)\,.\\
&g(m,\ell)=g(n,k)(n \rightarrow m; k \rightarrow \ell)\,,
\end{align}
\end{subequations}
By using the following integral representation of hypergeometric
type function $_2F_1(a,b;c;z)$ [18]
\begin{eqnarray}
_2F_1(a,b;c;z)=\,\frac{\Gamma(c)}{\Gamma(b)\Gamma(c-b)}\,\int_{0}^{1}
t^{b-1}\,(1-t)^{c-b-1}\,(1-tz)^{-a}\,dt\,,
\end{eqnarray}
and by setting the variable $z \rightarrow q$ and taking
$a=1+A+k+\ell$, $b=2a_{1}+n+m-k-\ell+1$, $c=1+b$, Eq. (20) gives
the normalization constant as
\begin{eqnarray}
A_{n}=\sqrt{\frac{q^{-2a_{1}-n-m+k+\ell-1}(-1)^{k+\ell-n-m}\Gamma(2a_{1}+n+m-k-\ell+2)}
{g(n,k)g(m,\ell)\Gamma(2a_{1}+n+m-k-\ell+1)\,_2F_{1}(a,b;c;q)}\,}\,.
\end{eqnarray}
\section{Creation and Annihilation Operators}
We want to obtain the following eigenvalue equations for creation,
$\hat{\Pi}_{+}$, and annihilation, $\hat{\Pi}_{-}$, operators
\begin{eqnarray}
\hat{\Pi}_{\mp}|n>=\pi_{\mp}|n>\,,
\end{eqnarray}
where the ket $|n>$ corresponds to the eigenfunctions $R_{n}(x)$.
Firstly, we study the effect of the differential operator
$\frac{d}{dx}$ on the eigenfunctions
\begin{eqnarray}
\frac{d}{dx}|n>=\left(\frac{a_{1}}{x}-\frac{1+A}{2(1-x)}\right)|n>+A_{n}x^{a_{1}}(1-x)^{(1+A)/2}\frac{d}{dx}
P_{n}^{(2a_{1},A)}(1-2x)\,,
\end{eqnarray}
With the help of the following relation [18]
\begin{eqnarray}
\frac{d}{dx}\left[P_{n}^{(\eta_{1},\eta_{2})}(x)\right]=\frac{1}{2}(n+\eta_{1}+\eta_{2}+1)
P_{n-1}^{(\eta_{1}+1,\eta_{2}+1)}(x)
\end{eqnarray}
we obtain the first derivative in the last term of Eq. (25) as
\begin{eqnarray}
\frac{d}{dx}\left[P_{n}^{(2a_{1},A)}(1-2x)\right]=-(n+2a_{1}+A+1)
P_{n-1}^{(2a_{1}+1,A+1)}(x)
\end{eqnarray}
Using the recursion relations for the Jacobi polynomials [18]
\begin{subequations}
\begin{align}
&P_{n}^{(\eta_{1},\eta_{2})}(x)=\frac{\eta_{1}+\eta_{2}+n+1}{\eta_{1}+n+1}P_{n}^{(\eta_{1}+1,\eta_{2})}(x)
-\frac{\eta_{2}+n}{\eta_{1}+n+1}P_{n}^{(\eta_{1}+1,\eta_{2}-1)}(x)\,,\\
&P_{n}^{(\eta_{1}+1,\eta_{2})}(x)=\frac{2}{2n+\eta_{1}+\eta_{2}+2}\,\frac{1}{1-x}\,
\left[(n+\eta_{1}+1)P_{n}^{(\eta_{1},\eta_{2})}(x)-(n+1)P_{n-1}^{(\eta_{1},\eta_{2})}(x)\right]\,,\\
&P_{n}^{(\eta_{1},\eta_{2}+1)}(x)=\frac{2}{2n+\eta_{1}+\eta_{2}+2}\,\frac{1}{1+x}\,
\left[(n+\eta_{2}+1)P_{n}^{(\eta_{1},\eta_{2})}(x)+(n+1)P_{n+1}^{(\eta_{1},\eta_{2})}(x)\right]\,,
\end{align}
\end{subequations}
Eq. (25) becomes
\begin{eqnarray}
\frac{d}{dx}|n>=\left(\frac{a_{1}-\kappa_{3}}{x}-\frac{\frac{1+A}{2}-\kappa_{2}}{1-x}\right)|n>
+\left(\frac{\kappa_{1}}{x}+\frac{\kappa_{1}}{1-x}\right)\frac{A_{n}}{A_{n-1}}|n-1>\,,
\end{eqnarray}
with
\begin{eqnarray}
\kappa_{1}=\frac{(n+A)(n+2a_{1})}{2n+2a_{1}+A}\,\,;\,\,\kappa_{2}=\frac{n(n+2a_{1})}{2n+2a_{1}+A}\,\,;\,\,
\kappa_{3}=\frac{n(n+A)}{2n+2a_{1}+A}\,,
\end{eqnarray}
Rewriting Eq. (29) in the standard form
\begin{eqnarray}
\left[x(1-x)\frac{d}{dx}+\left(a_{1}+\frac{1+A}{2}-\kappa_{2}-\kappa_{3}\right)x-a_{1}+\kappa_{3}\right]
\nonumber\\ \times
(\frac{2n+2a_{1}+A}{n+2a_{1}})|n>=(n+A)\frac{A_{n}}{A_{n-1}}|n-1>\,,
\end{eqnarray}
we obtain the annihilation operator
\begin{eqnarray}
\hat{\Pi}_{-}=\left[x(1-x)\frac{d}{dx}+\frac{1}{2}\,(2a_{1}+A-2(\kappa_{2}+\kappa_{3})+1)x-a_{1}+\kappa_{3}\right]
(\frac{2n+2a_{1}+A}{n+2a_{1}})\,,
\end{eqnarray}
with the eigenvalues
\begin{eqnarray}
\pi_{-}=(n+A)\frac{A_{n}}{A_{n-1}}\,.
\end{eqnarray}
which satisfy the eigenvalue equation in Eq. (24).

Following the same procedure and using the following relation [18]
\begin{eqnarray}
P_{n-1}^{(\eta_{1},\eta_{2})}(x)=P_{n}^{(\eta_{1},\eta_{2}-1)}(x)-P_{n}^{(\eta_{1}-1,\eta_{2})}(x)\,,
\end{eqnarray}
we immediately write the creation operator as
\begin{eqnarray}
\hat{\Pi}_{+}=\left[x(1-x)\frac{d}{dx}+\frac{1}{2}\,(2a_{1}+A+2(\kappa_{4}-\kappa_{5})+1)x-a_{1}-\kappa_{4}\right]
(\frac{2n+2a_{1}+A+2}{n+2a_{1}+A+1})\,,
\end{eqnarray}
where
\begin{eqnarray}
\kappa_{4}=\frac{(n+2a_{1}+A+1)(n+2a_{1}+1)}{2n+2a_{1}+A+2}\,\,;\,\,
\kappa_{5}=\frac{(n+2a_{1}+A+1)(n+A+1)}{2n+2a_{1}+A+2}\,,
\end{eqnarray}
with the eigenvalues
\begin{eqnarray}
\pi_{+}=(n+1)\frac{A_{n}}{A_{n+1}}\,.
\end{eqnarray}
Now, let us study the commutation relations of the operators
$\hat{\Pi}_{\mp}$ to present the Lie algebra related with the
two-term molecular potential. For this aim, we define the operator
\begin{eqnarray}
\hat{\Pi}_{0}=\hat{n}+\frac{1+A}{2}\,,
\end{eqnarray}
where $\hat{n}$ is the number operator satisfying
\begin{eqnarray}
\hat{n}|n>=n|n>\,,
\end{eqnarray}
So, the operator $\hat{\Pi}_{0}$ has the eigenvalues
\begin{eqnarray}
\pi_{0}=n+\frac{1+A}{2}\,.
\end{eqnarray}
By using Eqs. (32), (35), (38) and with the help of Eqs. (33),
(37) and (40), we obtain the commutators as
\begin{eqnarray}
\left[\hat{\Pi}_{-},\hat{\Pi}_{+}\right]=2\hat{\Pi}_{0}\,\,;\,\,\left[\hat{\Pi}_{0},\hat{\Pi}_{+}\right]=\hat{\Pi}_{+}
\,\,;\,\,\left[\hat{\Pi}_{-},\hat{\Pi}_{0}\right]=\hat{\Pi}_{-}\,,
\end{eqnarray}
which correspond to the Lie algebra associated with a SU(1,1)
group.

Finally, we could write the Casimir operator of the group by using
the above results as
\begin{eqnarray}
\hat{C}=-\hat{\Pi}_{-}\hat{\Pi}_{+}+\hat{\Pi}_{0}\left(\hat{\Pi}_{0}+1\right)=-\hat{\Pi}_{+}\hat{\Pi}_{-}+\hat{\Pi}_{0}\left(\hat{\Pi}_{0}-1\right)\,,
\end{eqnarray}
with the eigenvalue equation
\begin{eqnarray}
\hat{C}|n>=c_{0}|n>\,\,;\,\,\,
\end{eqnarray}
where
\begin{eqnarray}
c_{0}=c(c-1)\,\,;\,\,\,\,\,c=\frac{1+A}{2}\,.
\end{eqnarray}

\section{Conclusions}
We have computed the exact bound state energy eigenvalues and the
corresponding normalized eigenfunctions of the two-term diatomic
potential by using the NU method. Then we have obtained the
step-up and step-down operators of the potential and shown that
the ladder operators satisfy the Lie algebra of a SU(1,1) group.
The Casimir operator of the group is also obtained and the
eigenvalue equation is given.

\section{Acknowledgments}
This research was partially supported by the Scientific and
Technical Research Council of Turkey.

\appendix

\section{Nikiforov Uvarov Method}
The Schr\"{o}dinger equation can be transformed by using
appropriate coordinate transformation into following form
\begin{eqnarray}
\sigma^2(y)\frac{d^2\Psi(y)}{dy^2}+\sigma(y)\tilde{\tau}(y)\frac{d\Psi(y)}{dy}+\tilde{\sigma}(y)
\Psi(y)=0\,,
\end{eqnarray}
where $\sigma(y)$ and $\tilde{\sigma}(y)$ are polynomials, at
most, second degree, and $\tilde{\tau}(y)$ is a first-degree
polynomial. By using the separation of variables, the solution is
written as
\begin{eqnarray}
\Psi(y)=\psi(y)\varphi(y)\,,
\end{eqnarray}
which gives Eq. (A1) as a hypergeometric type equation [5]
\begin{eqnarray}
\frac{d^2\varphi(y)}{dy^2}+\frac{\tau(y)}{\sigma(y)}\frac{d\varphi(y)}{dy}+\frac{\lambda}{\sigma(y)}\,
\varphi(y)=0\,,
\end{eqnarray}
where $\psi(y)$ is defined by using the equation [5]
\begin{eqnarray}
\frac{1}{\psi(y)}\frac{d\psi(y)}{dy}=\frac{\pi(y)}{\sigma(y)}\,,
\end{eqnarray}
and the other part of the solution in Eq. (A2) is given by
\begin{eqnarray}
\varphi_{n}(y)=\frac{a_{n}}{\rho(y)}\frac{d^n}{dy^n}[\sigma^{n}(y)\rho(y)]\,,
\end{eqnarray}
where $a_{n}$ is a normalization constant and $\rho(y)$ is the
weight function and satisfies the following equation [5]
\begin{eqnarray}
\left[\sigma(y)\rho(y)\right]'=\tau(y)\rho(y)\,.
\end{eqnarray}
where prime denotes the derivative to coordinate.

The function $\pi(y)$ and the parameter $\lambda$ in the above
equation are defined as
\begin{eqnarray}
\pi(y)&=&\,\frac{1}{2}\,[\sigma'(y)-\tilde{\tau}(y)]
\pm\sqrt{\frac{1}{4}\left[\sigma'(y)-\tilde{\tau}(y)\right]^2
-\tilde{\sigma}(y)+k\sigma(y)\,}\,,\\
\lambda&=&k+\pi'(y)\,.
\end{eqnarray}
In the NU method, the square root in Eq. (A7) must be the square
of a polynomial, so the parameter $k$ can be determined. Thus, a
new eigenvalue equation becomes
\begin{eqnarray}
\lambda=\lambda_{n}=-n\tau'(y)-\frac{1}{2}\,n(n-1)\sigma''(y)\,.
\end{eqnarray}
where prime denotes the derivative and the derivative of the
function $\tau(y)=\tilde{\tau}(y)+2\pi(y)$ should be negative.


\newpage



\begin{thebibliography}{99}

\bibitem{ref1} J.~X.~Sun, Acta Phys. Sin. {\bf 48}, 1992 (1999).

\bibitem{ref2} C.~S.~Jia, J.~Y.~Wang, S.~He and L.~T.~Sun, J. Phys. A {\bf 33}, 6993 (2000).

\bibitem{ref3} F.~Benamira, L.~Guechi, S.~Mameri and M.~A.~Sadoun, J. Math. Phys. {\bf 48}, 032102 (2007).

\bibitem{ref4} A.~Arda and R.~Sever, J. Math. Chem. DOI:~10.1007/s10910-012-0011-0 (in press).

\bibitem{ref5} A.~F.~Nikiforov, and V.~B.~Uvarov, \textit{Special Functions of
Mathematical Physics }, (Birkh\"{a}user, Basel, 1988).

\bibitem{ref6} L.~Infeld and T.~E.~Hull, Rev. Mod. Phys. {\bf 23}, 21 (1951).

\bibitem{ref7} Y.~B.~Ding, J. Phys. A {\bf 20}, 6293 (1987).

\bibitem{ref8} E.~Schrödinger, Proc. R. Irish Acad. {\bf 47A}, 53 (1941).

\bibitem{ref9} S.~H.~Dong, App. Math. Lett. {\bf 16}, 199 (2003).

\bibitem{ref10} S.~H.~Dong, R.~Lemus and A.~Frank, Int. J. Quant. Chem. {\bf 86}, 433 (2002).

\bibitem{ref11} S.~H.~Dong and Z.~Q.~Ma, Am. J. Phys. {\bf 70}, 520 (2002).

\bibitem{ref12} S.~H.~Dong and Z.~Q.~Ma, Int. J. Mod. Phys. E {\bf 11}, 155 (2002).

\bibitem{ref13} S.~H.~Dong, G.~H.~Sun and M.~Lozada-Cassou, Phys. Lett. A {\bf 328}, 299 (2004).

\bibitem{ref14} S.~H.~Dong, Can. J. Phys. {\bf 80}, 129 (2002).

\bibitem{ref15} S.~H.~Dong, G.~H.~Sun and Y.~Tang, Int. J. Mod. Phys. E {\bf 12}, 809 (2003).

\bibitem{ref16} S.~H.~Dong, G.~H.~Sun and M.~Lozada-Cassou, Int. J. Mod. Phys. AE {\bf 20}, 5663 (2005).

\bibitem{ref17} S.~Flügge, \textit{Practical Quantum Mechnics I} (Springer Verlag, Berlin, Heidelberg, New York, 1971).

\bibitem{ref18} M.~Abramowitz, I.~A.~Stegun, \textit{Handbook of Mathematical
Functions with Formulas, Graphs, and Mathematical Tables} (Dover
Publications, New York, 1965)
\end{thebibliography}
\end{document}